\documentclass[conference]{IEEEtran}

\IEEEoverridecommandlockouts

\usepackage{amsmath,amsthm,relsize}
\usepackage{amsfonts, amssymb, cuted}
\usepackage{cleveref}
\usepackage{mathtools}
\usepackage[boxed]{algorithm}
\usepackage[noend]{algpseudocode}
\usepackage{cite}
\usepackage{xcolor}
\usepackage{soul}
\usepackage{graphicx}
\usepackage{tikz}
\usepackage{pgfplotstable}
\usepackage{pgfplots}
\usepackage{nicefrac}
\usepackage{enumitem}
\usepackage{support-caption}
\usepackage{subcaption}
\usepackage[font=footnotesize]{caption}
\usepackage{multirow}
\pgfplotsset{compat=1.15}
\usepackage{relsize}
\usetikzlibrary{patterns}
\usepackage{acro}
\usepackage{color,soul}

\usepackage{subcaption}           

\newcommand{\vbar}{\raisebox{.17ex}{\rule{.04em}{1.35ex}}}
\newcommand{\vbarind}{\raisebox{.01ex}{\rule{.04em}{1.1ex}}}
\newcommand{\R}{\ifmmode{\rm I}\hspace{-.2em}{\rm R} \else ${\rm I}\hspace{-.2em}{\rm R}$ \fi}
\newcommand{\T}{\ifmmode{\rm I}\hspace{-.2em}{\rm T} \else ${\rm I}\hspace{-.2em}{\rm T}$ \fi}
\newcommand{\N}{\ifmmode{\rm I}\hspace{-.2em}{\rm N} \else \mbox{${\rm I}\hspace{-.2em}{\rm N}$} \fi}
\newcommand{\B}{\ifmmode{\rm I}\hspace{-.2em}{\rm B} \else \mbox{${\rm I}\hspace{-.2em}{\rm B}$} \fi}
\newcommand{\Hil}{\ifmmode{\rm I}\hspace{-.2em}{\rm H} \else \mbox{${\rm I}\hspace{-.2em}{\rm H}$} \fi}
\newcommand{\C}{\ifmmode\hspace{.2em}\vbar\hspace{-.31em}{\rm C} \else \mbox{$\hspace{.2em}\vbar\hspace{-.31em}{\rm C}$} \fi}
\newcommand{\Cind}{\ifmmode\hspace{.2em}\vbarind\hspace{-.25em}{\rm C} \else \mbox{$\hspace{.2em}\vbarind\hspace{-.25em}{\rm C}$} \fi}
\newcommand{\Q}{\ifmmode\hspace{.2em}\vbar\hspace{-.31em}{\rm Q} \else \mbox{$\hspace{.2em}\vbar\hspace{-.31em}{\rm Q}$} \fi}
\newcommand{\Z}{\ifmmode{\rm Z}\hspace{-.28em}{\rm Z} \else ${\rm Z}\hspace{-.28em}{\rm Z}$ \fi}

\newtheorem{exmp}{Example}
\theoremstyle{definition}

\newcommand{\CA}[0]{{\mathcal{A}}}
\newcommand{\CB}[0]{{\mathcal{B}}}

\newcommand{\CS}[0]{{\mathcal{S}}}
\newcommand{\CT}[0]{{\mathcal{T}}}
\newcommand{\CU}[0]{{\mathcal{U}}}

\newcommand{\Ba}[0]{{\mathbf{a}}}

\newcommand{\Bc}[0]{{\mathbf{c}}}

\newcommand{\Bh}[0]{{\mathbf{h}}}

\newcommand{\Bp}[0]{{\mathbf{p}}}

\newcommand{\Bs}[0]{{\mathbf{s}}}

\newcommand{\Bu}[0]{{\mathbf{u}}}
\newcommand{\Bv}[0]{{\mathbf{v}}}
\newcommand{\Bw}[0]{{\mathbf{w}}}
\newcommand{\Bx}[0]{{\mathbf{x}}}

\newcommand{\BH}[0]{{\mathbf{H}}}

\newcommand{\BM}[0]{{\mathbf{M}}}

\newcommand{\SfL}[0]{{\mathsf{L}}}

\usepackage{titlesec}
\titlespacing\section{3pt}{6pt plus 4pt minus 2pt}{6pt plus 2pt minus 2pt}
\titlespacing\subsection{3pt}{4pt plus 4pt minus 2pt}{4pt plus 2pt minus 2pt}
\titlespacing\subsubsection{3pt}{3pt plus 4pt minus 2pt}{0pt plus 2pt minus 3pt}

\makeatletter
\newcommand\fs@betterruled{%
  \def\@fs@cfont{\bfseries}\let\@fs@capt\floatc@ruled
  \def\@fs@pre{\vspace*{8pt}\hrule height.8pt depth0pt \kern2pt}%
  \def\@fs@post{\kern2pt\hrule\relax}%
  \def\@fs@mid{\kern2pt\hrule\kern2pt}%
  \let\@fs@iftopcapt\iftrue}
\floatstyle{betterruled}
\restylefloat{algorithm}
\makeatother

\title{Collaborative Coded Caching for Partially Connected Networks}

\begin{document}

\author{\IEEEauthorblockN{Kagan Akcay\IEEEauthorrefmark{1}, Eleftherios Lampiris\IEEEauthorrefmark{1}, MohammadJavad Salehi\IEEEauthorrefmark{2}, and Giuseppe Caire\IEEEauthorrefmark{1}} \\
\IEEEauthorblockA{
    \IEEEauthorrefmark{1} Electrical Engineering and Computer Science Department, Technische Universit\"at Berlin, 10587 Berlin, Germany\\
    \IEEEauthorrefmark{2} Centre for Wireless Communications, University of Oulu, 90570 Oulu, Finland \\
    \textrm{kagan.akcay@tu-berlin.de \quad eleftherios.lampiris@tu-berlin.de \quad mohammadjavad.salehi@oulu.fi \quad caire@tu-berlin.de}
    }
}

\maketitle

\begin{abstract}
Coded caching leverages the differences in user cache memories to achieve gains that scale with the total cache size, alleviating network congestion due to high-quality content requests. Additionally, distributing transmitters over a wide area can mitigate the adverse effects of path loss. In this work, we consider a partially connected network where the channel between distributed transmitters (helpers) and users is modeled as a distributed multiple-input-multiple-output (MIMO) Gaussian broadcast channel. We propose a novel delivery scheme consisting of two phases: \emph{partitioning} and \emph{transmission}. In the partitioning phase, users with identical cache profiles are partitioned into the minimum number of sets, such that users within each set can successfully decode their desired message from a joint transmission enabled by MIMO precoding. To optimally partition the users, we employ the branch and bound method. In the transmission phase, each partition is treated as a single entity, and codewords are multicast to partitions with distinct cache profiles. The proposed delivery scheme is applicable to any partially connected network, and while the partitioning is optimal, the overall delivery scheme, including transmission, is heuristic. 
Interestingly, simulation results show that its performance closely approximates that of the fully connected optimal solution.
\end{abstract}
\section{Introduction}
Modern wireless telecommunications systems face an ever-increasing demand from users to deliver higher quality of service, faster data rates, and lower latency. As a result, telecommunication networks are under immense pressure to evolve and meet these expectations. Several factors contribute to the difficulty in meeting these demands, including: i) users being located close to the edge of the network, which results in lower data rates due to path loss and signal attenuation; ii) network congestion, where transmitters become overloaded as users request high-quality content, such as movies or real-time video streaming. 

To address these challenges, various technologies have been proposed to mitigate the impact of the limitations above. For example, to counteract the effects of path loss, one promising solution involves deploying multiple transmitters, such as WiFi routers, spread across a wide area~\cite{handover_networks}. This ensures that users remain within a reasonable distance from at least one transmitter, improving signal strength and reducing the likelihood of experiencing low data rates. 





Meanwhile, the demand for content-related applications has been steadily increasing, along with rising quality requirements. One promising technology that has the potential to alleviate these challenges is \emph{coded caching}~\cite{fundamental_caching}. With coded caching, users can pre-store some of the file contents during off-peak times to reduce network congestion and achieve gains that scale with the cumulative cache size of the network. Many works have built on the original coded caching work of~\cite{fundamental_caching}, e.g., for 
multi-server~\cite{multi-server}, Device-to-Device (D2D)~\cite{d2d_caching}, transmitter-side cache~\cite{cache-interference}, shared-cache~\cite{caire}, multi-access~\cite{multi_access_caching}, combinatorial~\cite{combinatorial_caching}, dynamic~\cite{dynamic_caching} networks, and ~\cite{physical_layer_caching,lefteris,parinello} combined multi-antenna/transmitter strategies with coded caching.

\begin{figure}[t]
        \centering
        \includegraphics[width = 0.60\columnwidth]{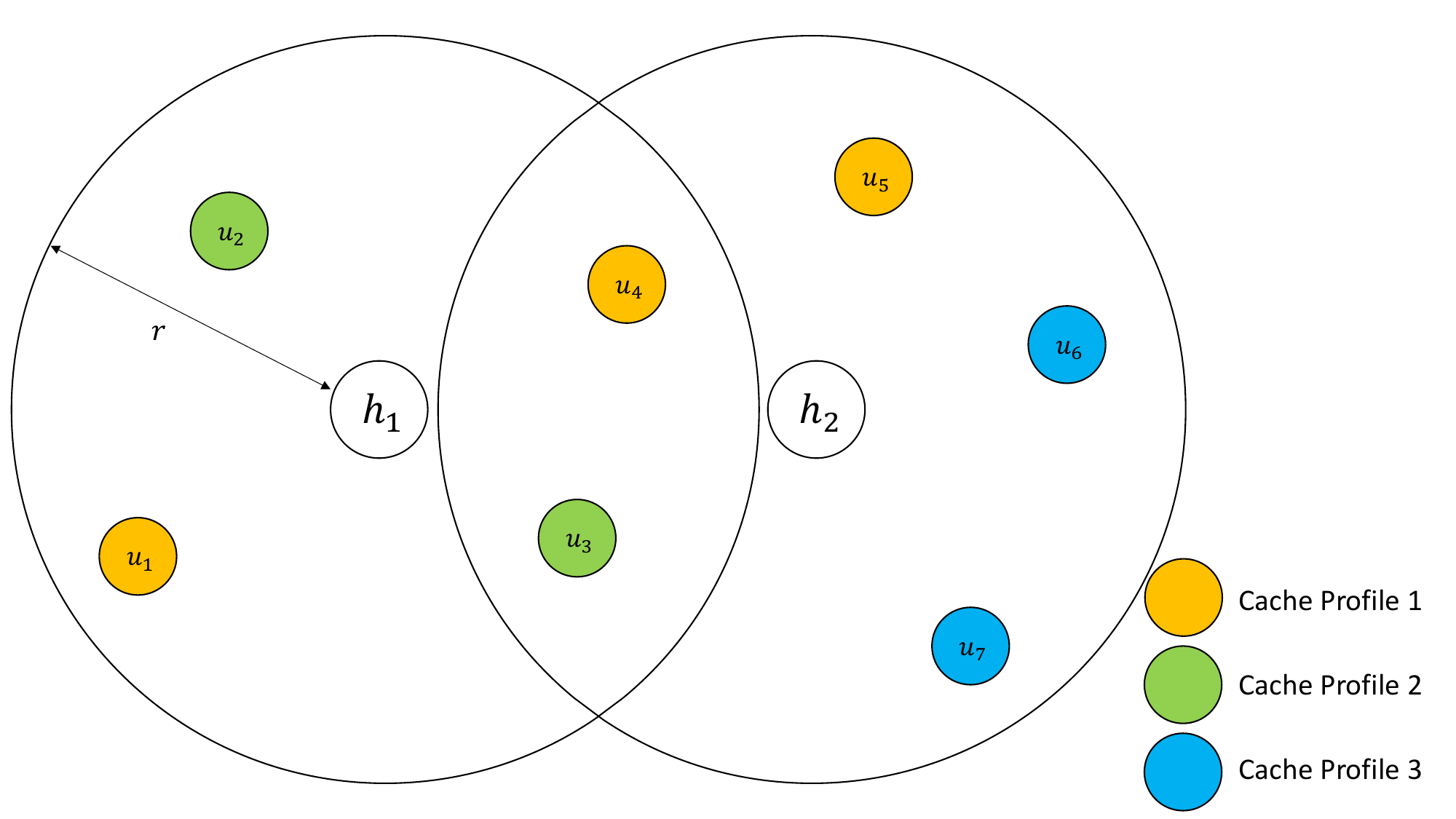} 
    \caption{A partially connected network with $E=2$ transmitters/helpers, $K=7$ users and $L = 3$ cache profiles.}
    \vspace{-6pt}
    \label{fig:network_model}
\end{figure}

Further, works \cite{mozhgan,kagan_1} have paired coded caching techniques with distributed transmitter strategies to relieve networks from the effects of path loss and congestion simultaneously. 
They considered a partially connected network where a server transmits data to multiple spatially distributed and cache-enabled users through several access points (APs), referred to as ``helpers". A collision-type interference model was used, where packets are lost if a user receives the superposition of concurrent packets from different helpers that exceeds a certain interference threshold. Further,~\cite{kagan_2} extended this model by incorporating multiple antennas on the transmitter side. In this work, we extend the collision interference model discussed in \cite{mozhgan, kagan_1, kagan_2} by considering a more fundamental scenario: the channel between all the helpers and users is modeled as a multi-antenna (MIMO) Gaussian broadcast channel. This allows all helpers to function as a distributed multiple-antenna (joint) transmitter. We focus on the standard setting in coded caching, where users request arbitrary files from the library. The goal of the coding scheme is to minimize the worst-case delivery time across all user demands. 


This paper builds upon the shared-cache model of~\cite{caire} to reduce subpacketization, where some users share the same cache placement. Inspired by~\cite{lefteris,parinello}, we leverage the coded caching gain to serve users with different cache profiles, and the spatial multiplexing gain of the transmitters to serve users with identical cache profiles simultaneously. Unlike~\cite{lefteris,parinello}, where each user is within the coverage area of all transmitters, we face the challenge that each user is only associated with a limited subset of transmitters. The fully connected network model is relevant in a single-cell scenario with a multi-antenna base station, where all users are within reach of all antennas. In contrast, in an extended network where helpers are spatially distributed, it is unrealistic to assume that each user can receive a sufficiently strong signal from all helpers. This motivates using the partially connected geometric model, as proposed in~\cite{mozhgan,kagan_1,kagan_2}, while also considering full cooperation among all helpers at the physical layer through MIMO precoding. In a fully connected network, any subset of users with the same cache profile, equal in number to the transmitters or antennas, can be served together, as in~\cite{lefteris,parinello}. However, in a partially connected network, the subset of users with the same cache profile that can be served together depends on the specific network topology.


To address this, we propose a novel delivery scheme that combines coded caching and spatial multiplexing to optimize data delivery in partially connected networks. The proposed scheme consists of two parts: \emph{partitioning} and \emph{transmission}. In the \emph{partitioning} phase, we exploit the spatial multiplexing gain of the transmitters to partition users with the same cache profile into the minimum number of sets, such that users within each set can successfully decode their desired message from a joint transmission. To the best of our knowledge, this \emph{partitioning} problem has not been previously studied. To solve it optimally, we employ the least cost branch and bound method, which is commonly used to solve combinatorial optimization problems~\cite{original_branch,lc_branch,comb_optimization}. 
In the \emph{transmission} phase, we leverage the coded caching gain by treating each user partition as a single entity. Codewords are then multicast to different partitions, each with a distinct cache profile, following the approach in~\cite{lefteris,parinello}.


We also conduct simulations to evaluate the performance of the proposed delivery scheme, comparing it with a greedy approximation where users with the same cache profiles are partitioned by greedily assigning them to transmitters. Additionally, we compare the proposed scheme to the 
scalable heuristic developed in~\cite{kagan_1} for the collision interference model  
and the delivery scheme proposed in~\cite{parinello}, which is optimal for fully connected networks under the shared-cache model when the number of users per cache profile exceeds the number of transmitters. Although our delivery scheme is designed for partially connected networks and, together with \emph{partitioning} and \emph{transmission}, is heuristic, simulation results demonstrate that it performs closely to the fully connected optimal solution.

Several works have investigated partially connected networks with coded caching, where the nonzero channels are modeled as AWGN. Notable examples include the Wyner channels in~\cite{wigger_wyner_interference,lefteris_wyner} and the linear networks in~\cite{partial_linear_networks,partial_linear_pda}, where each user is connected to consecutive transmitters. However, to the best of our knowledge, no study has explored partially connected networks with coded caching under general connectivity conditions. The proposed delivery scheme in this paper is designed to be applicable to any partially connected network.








\textit{Notation:} Vectors are represented by bold small letters, matrices by bold big letters, and sets by calligraphic letters. $\mathbb{C}$  denotes the space of complex numbers, $|\Bv|$ denotes the number of elements in vector $\Bv$, $\Bv[i]$ is the $i$th element of vector $\Bv$, $\BM_{i,j}$ is the element in the $i$th row and $j$th column of matrix $\BM$, and $\CA \backslash \CB$ denotes the set of elements of $\CA$ not in $\CB$. For integer $J$, $[J]$ represents the set $\{1,2,\cdots,J\}$. $\binom{n}{k}$ denotes binomial coefficient, and its value is zero if $n < k$.

\section{System Model}
\label{sec:sys_model}


In our model, 
a server is connected to $E$ single antenna helpers (transmitters) via error-free fronthaul links and serves the requests of $K$ cache-enabled users. We assume that users are interested in receiving files from a library of $N$ files and can store a fraction of this library in their cache memory. Each library file is $F$ bits, and each user can cache $MF$ bits corresponding to a fraction $\gamma=M/N$ of all the files. We assume that the fronthaul links have sufficiently high capacity, so they do not form the communication bottleneck and 
each helper has an \textit{effective transmission radius}, $r$, 
such that any links between a helper and users outside of this radius are equal to zero. In contrast, the links for users within the radius are nonzero. 


System operation consists of two phases: placement and delivery. 
The placement phase is done offline, e.g., when users are connected to a WiFi device, while the delivery phase commences when each user requests a file from the library of $N$ files.
During the placement phase, users' cache memories are filled with file contents. In the delivery phase, using a transmission strategy, the server aims to deliver the remaining file contents to the users via the helpers. The helpers are assumed to transmit simultaneously, and the message received at user $u_k$, $k\in[K]$, takes the following form: 

\begin{equation}
    y_k=\Bh_k^T \Bx + w_k
\end{equation}
where $\Bx\in \mathbb{C}^{E \times 1}$ denotes the vector containing all the signals transmitted by the $E$ helpers satisfying an average power constraint $\mathbb{E}(||\Bx||^2)\leq P$, $\Bh_k\in \mathbb{C}^{E \times 1}$ denotes the channel vector where $\Bh_k[i]$ is the channel coefficient between helper $e_i$ and user $u_k$ following some i.i.d continuous distribution (e.g. Gaussian) and is $0$ if user $u_k$ is outside of the transmission radius of helper $e_i$,
and $w_k$ represents the unit power AWGN noise at user $u_k$. We assume that perfect channel state information exists throughout the (active) nodes as in \cite{multi-server,cache-interference,lefteris,parinello}, the fading process is statistically symmetric across users where each nonzero link between a helper and a user has a capacity of the form $\log(SNR)+ o(\log(SNR))$, and 
the system operates in a very high power regime such that the metric of interest is the sum-Degrees of Freedom (DoF), i.e., the number of users that can be served simultaneously with a rate that grows as $\log(P)$ for asymptotically large $P$.  

A time slot corresponds to transmitting a single file from the library to a single user, and we want to minimize the worst-case delivery time T, where T denotes the number of time slots required to serve all the users when each user requests a distinct file from the library, given a cache placement. In this way, the sum-DoF becomes $d_{\sum}=\frac{K(1-\gamma)}{T}$, when each user caches $\gamma$ portion of each file. In the seminal coded caching paper~\cite{fundamental_caching}, each file in the library is split into $\binom{K}{K\gamma}$ sub-files such that the caches of the users are filled with collections of these subfiles where the cache of each user is unique. The users having different caches is significantly important since codewords can be multicast to multiple users using the differences in caches where the exact number of users to whom codewords can be multicast turns out to be $K\gamma+1$. However, since $\binom{K}{K\gamma}$ grows exponentially with $K$ in the placement scheme of~\cite{fundamental_caching}, it requires a more practical approach for finite-sized files.

\section{Description of the Scheme}
\label{sec:scheme}
In this paper, we employ the cache placement scheme of~\cite{caire}, where users are assigned to $L<K$ distinct \emph{cache profiles}, and the cache content of each user is determined by its assigned profile. The main advantage of this cache placement scheme is its reduced subpacketization requirement~\cite{finite_fractional_caching,finite_caching,low_subpacketization,graph_caching,pda_caching,hypergraph_caching}, 
and its practicality, as displayed in~~\cite{parinello,mozhgan,kagan_1,kagan_2}.


To employ the cache placement scheme of~\cite{caire}, we first let $Q$ denote the subpacketization constraint, i.e., the maximum number of subfiles we can divide each file into, and let $W^n$ denote a file in the library where $n\in[N]$. 
Assuming $\gamma L$ is an integer,\footnote{If $\gamma L$ is not an integer, the scheme can be modified by cache sharing between two schemes with  $\lfloor \gamma L \rfloor$ and $ \lceil \gamma L \rceil$~\cite{fundamental_caching}.} 
every file $W^n$ is partitioned into $\binom{L}{\gamma L}\leq{Q}$ equal-sized subfiles $W^n_{\CS}$, 
indexed by all possible subsets $\CS \subseteq [L]$ of size $|\CS| = \gamma L$. 
Let $\SfL(u_k) \in [L]$ denote the cache profile assigned to user $u_k$. If  $\SfL(u_k) = \ell$, 
user $u_k$ caches subfiles $W^n_{\CS}$ for all $\CS \ni \ell$, $n\in[N]$. 

The network of interest is partially connected. A user is within the transmission radius of one or multiple transmitters. With coded caching, we can serve multiple users with different cache profiles simultaneously. By utilizing the spatial multiplexing gain of the transmitters, we can also serve users with the same cache profile together. This idea is inspired by the schemes of~\cite{lefteris,parinello}. 
where their delivery schemes are stated in the following:

Let $\Bc_{\ell}$ denote the vector of user indices assigned to cache profile $\ell$ with size $|\Bc_{\ell}| = C_{\ell}$, 
and $d_{\Bc_{\ell}[j]}$ denote the index of the file requested by user $\Bc_{\ell}[j]$, for $j\in[C_{\ell}]$. For a fully connected network where 
there is a nonzero link between each helper (transmitter) and each user, and $C_{\ell}=E$ for each $\ell\in [L]$, in~\cite{lefteris} 
the following vector

\begin{equation}
\label{eq:lefteris_transmission}
    \Bx(\CT)=\sum_{\ell \in \CT} \BH_{(\Bc_{\ell})}^{-1}\Bw_{(\Bc_{\ell})}
\end{equation}
is transmitted for each subset $\CT \subseteq [L]$ of size $|\CT|=\gamma L+1$ to satisfy all the user requests where $\BH_{(\Bc_{\ell})}^{-1}$ denotes the inverse of the channel matrix between the $E$ transmitters and the $C_{\ell}=E$ users with cache profile $\ell$, and $\Bw_{(\Bc_{\ell})}$ denotes the column vector where its $j$th element is   $W^{d_{\Bc_{\ell}[j]}}_{\CT \backslash \{ \ell \}}$, for $j\in[C_{\ell}]$. So, in each transmission, $E (\gamma L+1)$ users are served simultaneously. After~\eqref{eq:lefteris_transmission} is transmitted, user $u_{k'}$ with cache profile $\ell'$ where $\ell'\in{\CT}$ can remove  $\sum_{\ell \in \CT \backslash \{\ell'\}}\Bh_{k'}^T\BH_{(\Bc_{\ell})}^{-1}\Bw_{(\Bc_{\ell})}$ from its received message since $\ell'\in{\CT \backslash \{\ell\}}$ and $W^{d_{\Bc_{\ell}[j]}}_{\CT \backslash \{ \ell \}}$ is in the cache of user $u_{k'}$, for every $\ell \in \CT \backslash \{\ell'\}$ and every $j\in[C_{\ell}]$. User $u_k{'}$ then can decode its desired subfile $W^{d_{\Bc_{\ell'}[j']}}_{\CT \backslash \{ \ell' \}}$, where $j'$ corresponds to the index of $u_{k'}$ in the vector of users with cache profile $\ell'$, since  $\Bh_{k'}^T\BH_{(\Bc_{\ell'})}^{-1}\Bw_{(\Bc_{\ell'})}=W^{d_{\Bc_{\ell'}[j']}}_{\CT \backslash \{ \ell' \}}$. In this way, each user with cache profile $\ell\in{\CT}$ can decode a subfile from each transmission such that after~\eqref{eq:lefteris_transmission} is transmitted for each subset $\CT \subseteq [L]$ of size $|\CT|=\gamma L+1$, each user can decode their requested file.

For a fully connected network where there is a nonzero link between each helper (transmitter) and each user with $C_\ell \geq E$ for each $\ell\in [L]$,~\cite{parinello} further divides each subfile $W^n_{\CS}$ into $E$ equal-sized smaller subfiles $W^n_{\CS,i}$, for $i\in[E]$. User requests are satisfied in $\max_{\ell}{C_{\ell}}:=C$ rounds. Let $\Bc_\ell^E:=[\Bc_\ell,\Bc_\ell,...,\Bc_\ell]$ denote the $E$-fold concatenated vector of $\Bc_\ell$ for $\ell\in[L]$. In each round $o$, $o\in [C]$, users $\Bc_\ell^E[(o-1)*E+i]$ for each $i\in[E]$ and each $\ell\in[L]$ are served, if the element of the vector is nonempty. Notice that for given $o$ and $\ell$, $\Bc_\ell^E[(o-1)*E+i]$ is either empty or nonempty for all $i\in[E]$. For given $\CT \subseteq [L]$ of size $|\CT|=t+1$, let $\CT^o$ denote the set $\CT$ excluding the cache profiles $\ell$ where $\Bc_\ell^E[(o-1)*E+i]$ is empty for all $i\in[R]$ in round $o$. Then in round $o$, for each subset $\CT \subseteq [L]$ of size $|\CT|=\gamma L+1$ where $\CT^o$ is nonempty, the following vector is transmitted  

\begin{equation}
\label{eq:parinello_transmission}
    \Bx(\CT^o)=\sum_{l \in \CT^o} \BH_{(\Bc_{\ell}^R(o))}^{-1}\Bw_{(\Bc_{\ell}^R(o))}
\end{equation}
where $\BH_{(\Bc_{\ell}^E(o))}^{-1}$ denotes the inverse of the channel matrix between the $E$ transmitters and the $E$ users with cache profile $\ell$ in round $o$, and $\Bw_{(\Bc_{\ell}^E(o))}$ denotes the column vector where its $i$th element is   $W^{d_{\Bc_{\ell}^E[(o-1)*E+i]}}_{\CT \backslash \{ \ell \},(o,i)}$ for $i\in[E]$, $(o,i)$ corresponds to the next subindex of the subfile requested by user $\Bc_{\ell}^E[(o-1)*E+i]$. Decoding follows similarly to the decoding for the transmission in~\eqref{eq:lefteris_transmission}.

\cite{lefteris,parinello} consider a fully connected network and assume that the channels between each set of $E$ users and the $E$ transmitters are invertible. In this way, they can serve any $E$ users with the same cache profile using MIMO precoding, and their delivery schemes are based on this simple idea. However, for a partially connected network with possibly zero links between some transmitters and some users, as considered in this paper, this can not be assumed so that the schemes in~\cite{lefteris,parinello} are not applicable. The problem is assigning users to transmitters. Since not every user can be assigned to each helper, if a user has multiple helpers that can be assigned to, then which helper the user is assigned to may affect the performance of the delivery scheme. 
In this paper, we propose a delivery scheme with two parts: \emph{partitioning} and \emph{transmission}. In \emph{partitioning}, users corresponding to each cache profile are partitioned into the minimum number of sets where each user in a set can successfully decode its desired message from the joint transmission dedicated to the set. After users are partitioned accordingly, in \emph{transmission}, each partition of users is treated as a single user, and codewords are multicast to different partitions with different cache profiles, as in~\cite{lefteris,parinello}.
\subsection{Partitioning}
\label{sec:partition}
\begin{figure}[t]
        \centering
        \vspace{-16pt}
        \includegraphics[width = 0.7\columnwidth]{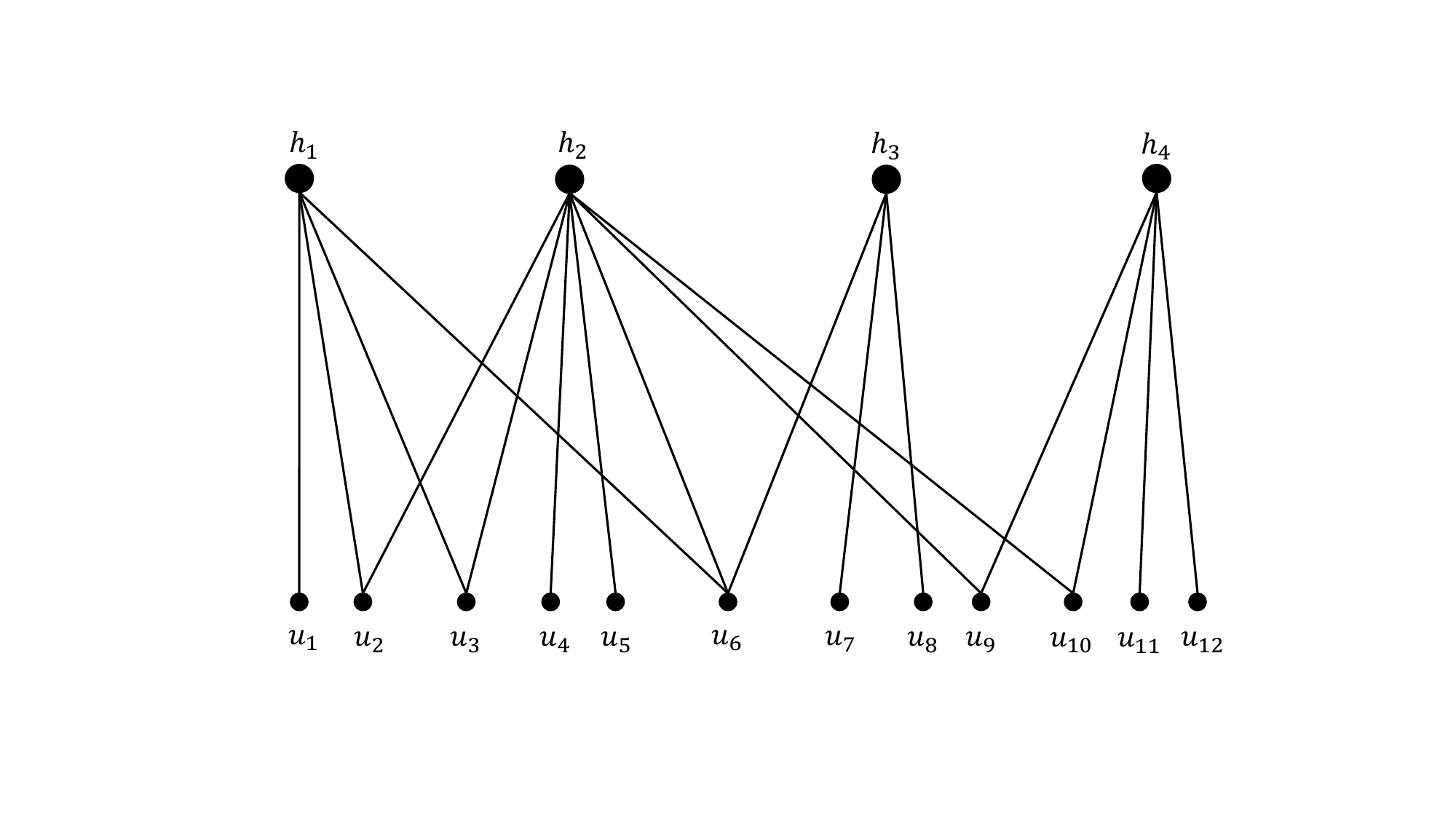} 
       \vspace{-16pt}
    \caption{Example subnetwork where each user has the same cache profile.}
    \vspace{-7.8pt}
    \label{fig:subnetwork}
\end{figure}
First, we split\footnote{When we split the network, we only split the users and their nonzero links to the helpers. We don't split the helpers.} the network into $L$ different sub-networks $\CT_{\ell}$. Each $\CT_{\ell}$ consists of $C_{\ell}$ users, each having cache profile $\ell$ (see Figure~\ref{fig:subnetwork}). 
Our aim is to partition the users in $\CT_{\ell}$ into the minimum number of different sets, where each user in a set can successfully decode its desired message from the joint transmission dedicated to the set. Notice that for a fully connected network, if every square submatrix $\textbf{H}^{E \times E}$ of the channel matrix of $\CT_{\ell}$ is invertible as assumed in~\cite{lefteris,parinello}, the minimum number of such sets as defined above is $\lceil\frac{C_{\ell}}{E}\rceil$, since the sum-DoF is equal to the number of helpers $E$ and every $E$ number of users can be served together by a joint transmission. It is easy to see that for a partially connected sub-network $\CT_{\ell}$, the minimum number of such sets will be lower bounded by $\lceil\frac{C_{\ell}}{E}\rceil$. If one considers the helpers, users with the same cache profile, and the nonzero links between them together as a bipartite graph (see Figure~\ref{fig:subnetwork}), the set of disjoint nonzero links can be seen as a matching~\cite{graph_theory}. It is well known~\cite{mahdi_invertible} that the rank of a matrix $\BH$ formed by identically $0$ elements and other elements drawn independently from an i.i.d continuous distribution (e.g., i.i.d. Gaussian) is with probability $1$ equal to the maximum matching in the bipartite graph formed by row indices, and column indices, where a row index $i$ and column index $j$ have an edge $(i,j)$ if $\BH_{i,j}$ is not identically zero. 
This implies that for any $D\leq E$ number of helpers and $D$ number of users, if there are disjoint nonzero links between $D$ helper-user pairs, for instance, between $e_1$ and $u_1$, $e_2$ and $u_2$,..., $e_D$ and $u_D$ where all $e_d$ and $u_d$ are distinct, all the $D$ users 
can decode their message from a joint transmission. 
  


So, the problem is reduced to finding disjoint nonzero links between helper-user pairs.  For simplicity, we will denote the partitions by only the user indexes. For instance, when we say $3-4-1-7$, this implies that there are nonzero links between $e_1$ and $u_3$, $e_2$ and $u_4$, $e_3$ and $u_1$, and $e_4$ and $u_7$ such that $u_3$, $u_4$, $u_1$ and $u_7$ can be served together. And if there is $0$ in a partition, the corresponding helper isn't assigned to a user.

\begin{exmp}
\label{ex:new_transmission}
Consider the subnetwork in Figure~\ref{fig:subnetwork}. The channel matrix $\BH$ between the $4$ helpers and users $u_1$, $u_2$, $u_6$ and $u_9$ can be written as  $\BH=\left[\begin{smallmatrix}
\Bh_{1}[1] & 0 & 0 & 0\\
\Bh_{2}[1] & \Bh_{2}[2] & 0 & 0 \\
\Bh_{6}[1] & \Bh_{6}[2] & \Bh_{6}[3] & 0 \\
0 & \Bh_{9}[2] & 0 & h_{9}[4]
\end{smallmatrix}\right].$
Notice that the nonzero elements in the diagonal constitute a matching. Assume $u_1$, $u_2$, $u_6$ and $u_9$ request the messages $X_1$, $X_2$, $X_3$, and $X_4$, respectively. Helpers can transmit $\Bx=[X_1,X_2-X_1\frac{\Bh_2[1]}{\Bh_2[2]},X_3-X_2\frac{\Bh_6[2]}{\Bh_6[3]}+X_1\frac{\Bh_2[1]\Bh_6[2]-\Bh_2[2]\Bh_6[1]}{\Bh_2[2]\Bh_6[3]},X_4-X_2\frac{\Bh_9[2]}{\Bh_9[4]}+X_1\frac{\Bh_2[1]\Bh_9[2]}{\Bh_2[2]\Bh_9[4]}]^T$
such that 
$\BH\Bx=[X_1\Bh_1[1],X_2\Bh_2[2],X_3\Bh_6[3],X_4\Bh_9[4]]^T
$, 
and $u_1$, $u_2$, $u_6$ and $u_9$ can decode $X_1$, $X_2$, $X_3$ and $X_4$, respectively. Notice that the partition $1-2-6-9$ is constructed by greedily assigning to each helper the first free user to which it has a nonzero link. It is easy to see that by continuing to assign the helpers to users greedily, one has the following partitions: $3-4-7-10$, $0-5-8-11$, and $0-0-0-12$. In this way, we have $4$ partitions. However, $\frac{C_{\ell}}{E}=\frac{12}{4}=3$, so the question is: Can we put all the users into less than $4$ partitions? 

\end{exmp}

The idea is this: We first split the users in each $\CT_{\ell}$ into two groups: the first group consists of users who can be served by only one helper, i.e., the users with only one nonzero link connected to a helper; the second group consists of users who can be served by more than one helper, i.e., the users with multiple nonzero links connected to distinct helpers. For instance, for the subnetwork in Figure~\ref{fig:subnetwork}, the first group consists of users $u_1$, $u_4$, $u_5$, $u_7$, $u_8$, $u_{11}$ and $u_{12}$, and the second group consists of $u_2$, $u_3$, $u_6$, $u_9$ and $u_{10}$. We also represent the user groups by two tables: the first group by $\CU_1$ and the second by $\CU_2$. In each table, the columns correspond to the helpers, and the rows correspond to the users. For $\CU_1$, if there is a nonzero link between a user $u_k$ and a helper $e_i$, the $i$-th column of the first empty row of $\CU_1$ is filled by the user's index, $k$. For $\CU_2$, if a user has nonzero links connected to multiple helpers $e_i$, each $i$-th column of the first empty row is filled by $k$. The first and second user groups in the subnetwork in Figure~\ref{fig:subnetwork} are represented by Table~\ref{tab:U_1} and Table~\ref{tab:U_2}, respectively.

    \begin{table}[t]
    \parbox{.45\linewidth}{
    \centering
    \begin{tabular}{|c|c|c|c|}
    \hline
           1 & 4 & 7 & 11                             \\
         \hline
         & 5 & 8 & 12  \\
         \hline
    \end{tabular}
    \caption{$\CU_1$ 
    }
    \label{tab:U_1}
    }
        \hfill
        \parbox{.45\linewidth}{
    \centering
    \begin{tabular}{|c|c|c|c|}
    \hline
           2 & 2 &  &                              \\
         \hline
          3 & 3 & &  \\
         \hline
         6 & 6 & 6 & \\
         \hline
         & 9 & & 9 \\
         \hline
         & 10 & & 10 \\
         \hline
    \end{tabular}
    \caption{$\CU_2$
    }
    \label{tab:U_2}
    }
    \end{table}

Every partition consists of disjoint nonzero links between helper-user pairs such that each different helper can only be assigned to a distinct user, so in every partition, a maximum of one user can be chosen from each different column corresponding to a different helper, but also a maximum of one user from the two same columns of tables $\CU_1$ and $\CU_2$, corresponding to the same helper. In addition, if a user is chosen from $\CU_2$, all the corresponding row of that user is erased. For instance, it can easily be verified from Tables~\ref{tab:U_1} and~\ref{tab:U_2} that $1-5-7-12$, $2-3-6-9$ and $6-4-8-9$ are all partitions. It is also easy to see that every partition can be constructed by this set of rules from Tables~\ref{tab:U_1} and~\ref{tab:U_2}.

Recall that our aim is to put users into the minimum number of partitions. This means that each user must be in only one partition, and no user is left out. Every user from table $\CU_1$ can be assigned to only one helper. However, every user from table $\CU_2$ can be assigned to multiple helpers, and the set of helpers a user can be assigned to can be distinct. Also, the number of users in columns of table $\CU_1$ can be different. So, depending on which helpers the users in $\CU_2$ are assigned to may affect the number of partitions. 

First, we calculate the number of users in each column of table $\CU_1$ and put the users in each row of $\CU_1$ into partitions. Then, we use the least cost branch and bound method to 
assign the users in table $\CU_2$ to helpers to be put into partitions accordingly and minimize the number of partitions. In least cost branch and bound, an objective function $f$ whose argument takes values in a finite set is minimized (or maximized), where the minimization is done without checking all the possible solutions. 

The algorithm consists of two main steps: \emph{Branch} and \emph{Bound}. In \emph{Branch}, a given subset of the possible solutions is split into at least two nonempty subsets, and in \emph{Bound}, a lower bound (cost) $B$ on the values $f$ takes in a subset of solutions is computed. \emph{Branch} and \emph{Bound} are performed iteratively to find the optimal solution. First, an upper bound $f^{max}$ on the values $f$ takes is computed heuristically, or one sets $f^{max}=\infty$. If in a \emph{Bound} step $B\geq f^{max}$, then the corresponding subset of solutions is discarded. Among the different sets of subsets of possible solutions, the set with the smallest lower bound is \emph{branched} first. 

To apply the least cost branch and bound method to the ``partitioning" problem, we first set $f^{max}=\infty$. 
In each branch step, we split the subsets according to the multiple helpers to which the considered user can be assigned, and in each bound step, $B$ corresponds to the number of partitions at hand after the last considered user is assigned to its respective helper.

The pseudo-code of the least cost branch and bound algorithm is provided in Algorithm~\ref{alg:branch_and_bound}, where we have used the following notations: 
1) $\Bu$ denotes the vector of users from $\CU_2$ ordered according to their row index;
2) $\Bs$ denotes the strategy vector corresponding to the helpers the users in $\Bu$ assigned to such that $\Bs[j]$ is the helper assigned to user $\Bu[j]$; 
3) for given $\Bs$, $\Bs_u$ denotes the vector where each of its element $\Bs_u[i]$ corresponds to the number of users helper $h_i$ is assigned to; 
4) $\CS$ denotes the set of strategies $\Bs$. Initially, $\Bs$ and $\CS$ are empty, $\Bs_u[i]$ is the number of users in the $i$th column of $\CU_1$. We have also used some auxiliary functions: 
1) $\textsc{Candid}(\Bu[j])$ gives the vector of helpers that can be assigned to $\Bu[j]$; 
2) $\textsc{prev}(\Bs)$ erases the helper in the last element from $\Bs$; 
3) $\textsc{find}(\CS,\min_{\Bs\in{\CS}}B_{\Bs})$ finds a state $\Bs\in{\CS}$ with largest $|\Bs|$ that achieves $\min_{\Bs\in{\CS}}B_{\Bs}$. Finally, in line $14$ of Algorithm~\ref{alg:branch_and_bound}, $arg(.)$ gives the helper with the lowest index if there are multiple options.

\begin{algorithm}[h]
\caption{Least Cost Branch and Bound}
\label{alg:branch_and_bound}
\begin{algorithmic}[1]
\State $\textsc{Initialize}$
 \ForAll{$h_i\in \textsc{candid}(\Bu[j])$}
  \State $\Bs' \gets [\Bs, h_i]$
  \State $\Bs'_u[i] \gets \Bs_{u}[i]+1$
  \State $B_{\Bs'} \gets \max_{i^*\in[R]}\Bs'_u[i^*]$
 \State $\CS \gets \lbrace\CS,\Bs'\rbrace$
   \State $B_i \gets B_{\Bs'}$
  \EndFor
 \State $\CS \gets \CS \backslash \textsc{Prev}(\Bs')$
\State $B \gets \min_{h_i\in \textsc{candid}(\Bu[j])}B_i$
 \If{$B> \min_{\Bs\in{\CS}}B_{\Bs}$}
 \State $\Bs \gets \textsc{find}(\CS,\min_{\Bs\in{\CS}}B_{\Bs})$
 \If{$|\Bs|=|\Bu|$}
 \State Go to Line 21
 \EndIf
 \State $j \gets |\Bs|+1$
 \State Go to Line 2
 \EndIf
 \State $\Bs \gets [\Bs,arg(\min_{h_i\in \textsc{candid}(\Bu[j])}B_i)]$
 \State $j \gets j+1$
 \If{$j<|\Bu|$}
 \State Go to Line 2
 \Else
 \State Return $\Bs$, $B_{\Bs}$
 \EndIf
\end{algorithmic}
\end{algorithm}

\begin{figure}[t]
        \centering
        \includegraphics[width = 0.75\columnwidth]{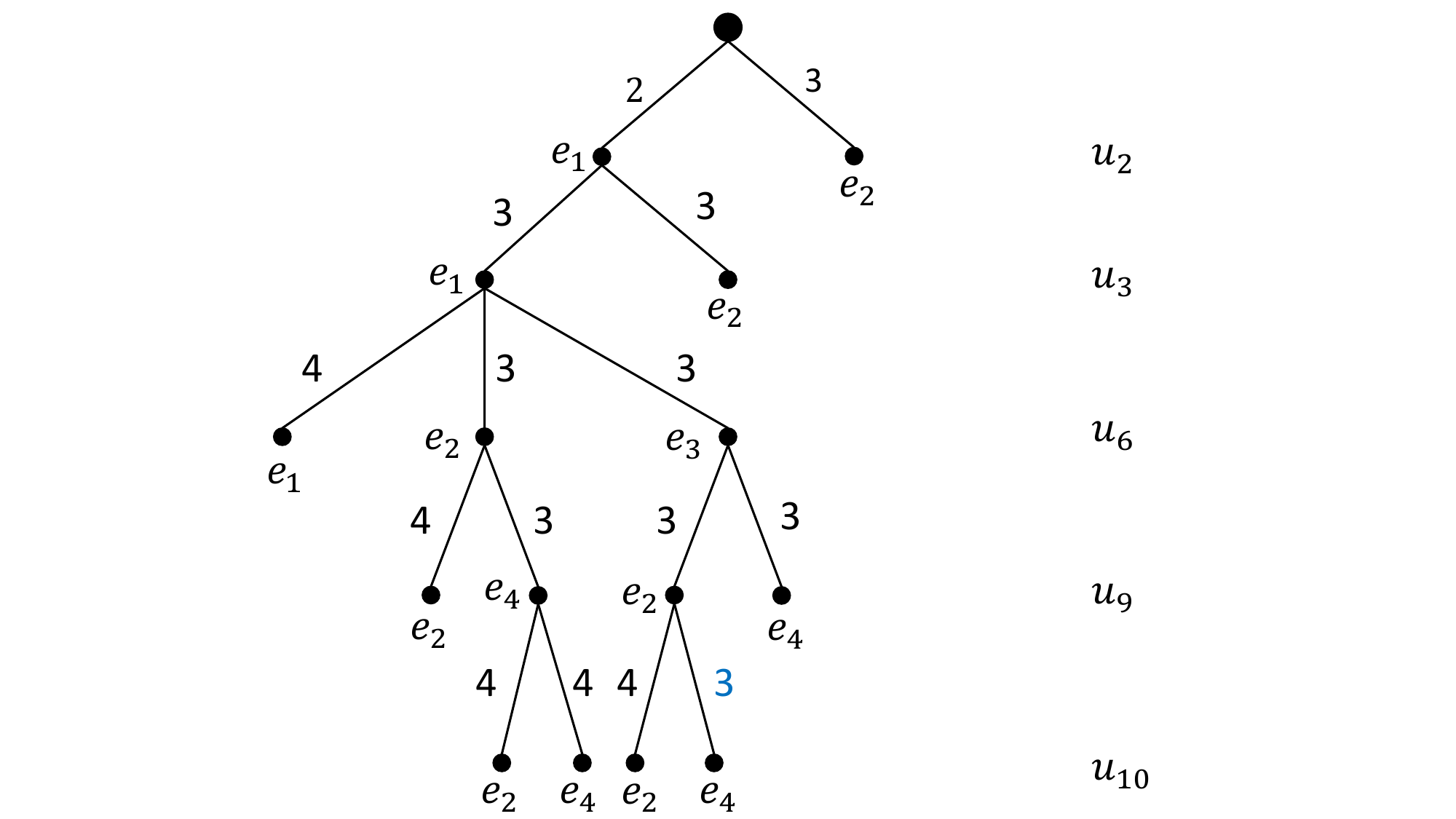} 
    \caption{Least cost branch and bound solution of the subnetwork in Figure~\ref{fig:subnetwork}. Each vertex corresponds to assigning a user to a helper, while an edge corresponds to the number of partitions after such an assignment.}
    \vspace{-4.5pt}
    \label{fig:branchandbound}
\end{figure}

Figure~\ref{fig:branchandbound} shows the graphical representation of the least cost branch and bound solution of the subnetwork in Figure~\ref{fig:subnetwork}. It can be seen from Table~\ref{tab:U_1} and Figure~\ref{fig:branchandbound} that the minimum number of partitions is $3$, where the partitions can be constructed as $1-4-7-11$, $2-5-8-12$, and $3-9-6-10$.
\subsection{Transmission}
\begin{figure*}
\begin{minipage}[t]{0.3\textwidth}
  \includegraphics[width=0.85\columnwidth]{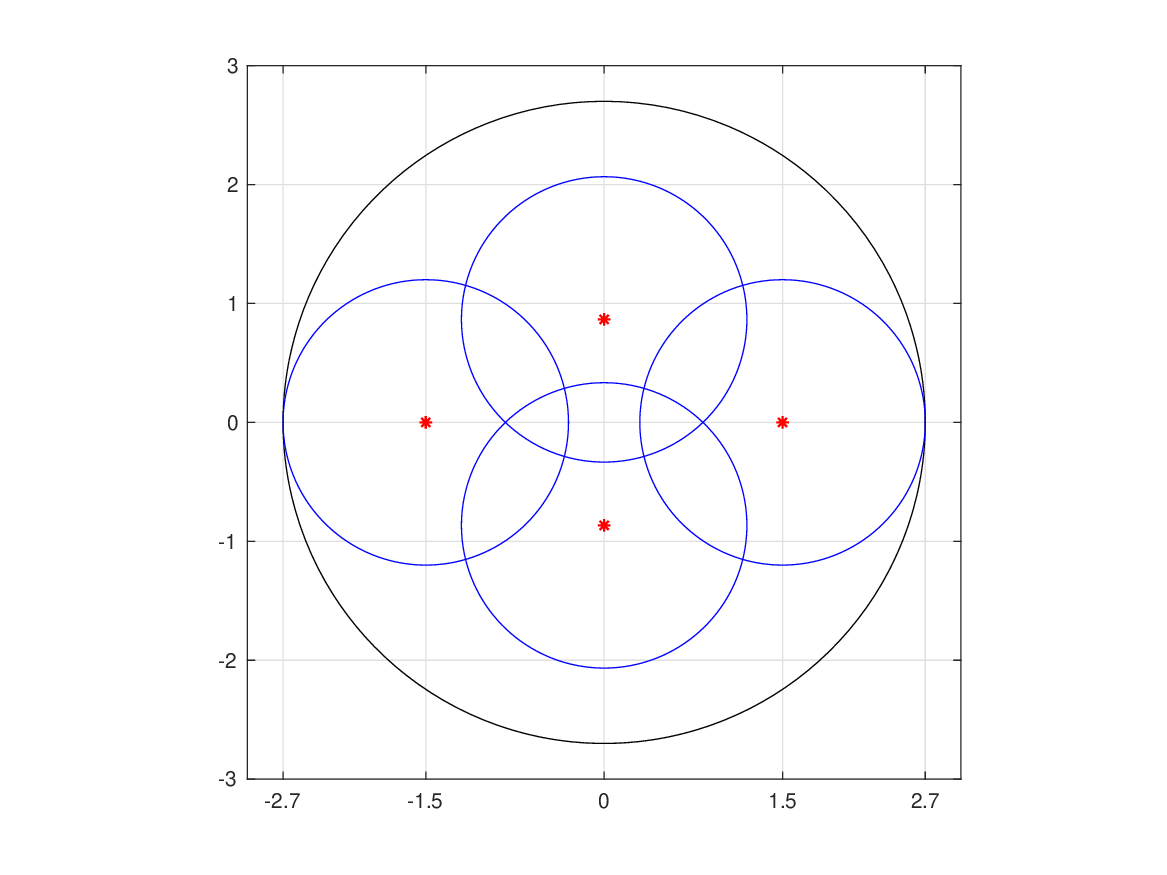}
  \caption{Hexagonal Placement of the Helpers, $r=1.2$, $r_u=2.7$.}
  \label{fig:hexagonal lattice}
\end{minipage}%
\hfill 
\begin{minipage}[t]{0.3\textwidth}
        
    \resizebox{0.8\columnwidth}{!}{%
    
    \begin{tikzpicture}

    \begin{axis}
    [
    axis lines = left,
    xlabel = \smaller {L},
    ylabel = \smaller {Sum-DoF},
    xmin=0, 
    ymin=0,
    xmax=42,
    ylabel near ticks,
    legend pos = south east,
    tick label style={font=\smaller},
    grid=both,
    major grid style={line width=.2pt,draw=gray!30},
    ]
    
    
    
    \addplot[black,mark=*,error bars/.cd,y dir=both,y explicit]
    coordinates{(1,3.024023809523809)+-(0,0.270928015306122)(10,4.976034075681626)+-(0,0.153929865604270)(20,6.876462639042871)+-(0,0.187187124057940)(30,8.885770733723222)+-(0,0.282449892088232)(40,10.670728915639783)+-(0,0.387947930885660)};
    \addlegendentry{\tiny Branch and Bound}
    \addplot[blue,mark=*,error bars/.cd,y dir=both,y explicit]
   coordinates{(1,2.840833333333334)+-(0,0.208581561791383)(10,4.692944572548156)+-(0,0.114811124198424)(20,6.521961684280152)+-(0,0.164950840492031)(30,8.401203243460596)+-(0,0.246965829901548)(40,10.146634423747184)+-(0,0.314207483495843)};
    \addlegendentry{\tiny Greedy}
    \addplot[red,mark=*,error bars/.cd,y dir=both,y explicit]
   coordinates{(1,2.413087301587302)+-(0,0.226977528911565)(10,4.301429163423493)+-(0,0.088901664533280)(20,5.660655773513193)+-(0,0.099867105376349)(30,7.054375746925614)+-(0,0.160497278208006)(40,8.201167614516347)+-(0,0.214764018447595)};
    \addlegendentry{\tiny Scalable Heuristic~\cite{kagan_1}}

    \end{axis}

    \end{tikzpicture}
    }
        \caption{Sum-DoF for different values of $L$, with $H=4$, $u/L=\frac{4}{(1.2)^2\pi}$, $\gamma=0.1$, $r=1.2$.}
     \label{fig:FixedU/L}
\end{minipage}%
\hfill
\begin{minipage}[t]{0.3\textwidth}
    \resizebox{0.8\columnwidth}{!}{%
    
    \begin{tikzpicture}

    \begin{axis}
    [
    axis lines = left,
    xlabel = \smaller {r},
    ylabel = \smaller {Sum-DoF},
    xmin=0, 
    ymin=0,
    xmax=4.5,
    ymax=7,
    ylabel near ticks,
    legend pos = south west, 
    tick label style={font=\smaller},
    grid=both,
    major grid style={line width=.2pt,draw=gray!30},
    ]
    
    
    
    \addplot[black,mark=*,error bars/.cd,y dir=both,y explicit]
    coordinates{(1.2,3.958455542891724)+-(0,0.161705876466675)(2.2,5.516825706916485)+-(0,0.150181464703537)(3.2,5.705196625932543)+-(0,0.058701554655324)(4.2,5.705196625932543)+-(0,0.058701554655324)};
    \addlegendentry{\tiny Branch and Bound}
    \addplot[blue,mark=*,error bars/.cd,y dir=both,y explicit]
   coordinates{(1.2,3.749893717702260)+-(0,0.166892875172209)(2.2,5.091953940780936)+-(0,0.176539469580207)(3.2,5.580997259801683)+-(0,0.079633053716999)(4.2,5.705196625932543)+-(0,0.058701554655324)};
    \addlegendentry{\tiny Greedy}
    \addplot[red,mark=*,error bars/.cd,y dir=both,y explicit]
   coordinates{(1.2,3.574297186407217)+-(0,0.049938517820173)(2.2,2.416119356048990)+-(0,0.026030889881811)(3.2,1.816131814877606)+-(0,0.006074833525162)(4.2,1.699419923606816)+-(0,0.003024200678699)};
    \addlegendentry{\tiny Scalable Heuristic~\cite{kagan_1}}
    \addplot[orange,thick]
    coordinates{(0,6.797679694427262)(4.5,6.797679694427262)};
    \addlegendentry{\tiny Upper Bound~\cite{parinello}}

    \end{axis}

    \end{tikzpicture}
    }
       \caption{Sum-DoF for different values of $r$, with $H=4$, $L=10$, $u=\frac{12}{(1.2)^2\pi}$, $\gamma=0.1$.}
     \label{fig:ChangeR}
\end{minipage}%
\vspace{-15pt}
\end{figure*}
Let  $G_\ell$ denote the number of partitions in $\CT_{\ell}$. For given $g\in[G_\ell]$, let  $\Ba_{\ell,g}$ denote the vector corresponding to the helpers the users are assigned to in the $g$-th partition in $\CT_{\ell}$ constructed in Section~\ref{sec:partition}, let $\Bp_{\ell,g}$ denote the users in the partition where $U_{\ell,g}$ denotes the number of users in it.  
For instance, for given $g$ and $l$, if the corresponding partition is $0-2-0-1$, then $\Ba_{\ell,g}=[e_2,e_4]$ and $\Bp_{\ell,g}=[u_2,u_1]$.  
Let also $\max_\ell{G_\ell}:=G$. 
There will be a total of $G$ rounds, where in each round $g\in[G]$, the users in partitions $\Bp_{\ell,g}$ for each $\ell\in[L]$ are served if the corresponding partition is not empty.   
For given $\CT_{j} \subseteq [L]$ of size $|\CT_{j}|=\gamma L+1$, $j\in [\binom{L}{\gamma L+1}]$, let $\CT^g_{j}$ denote the set $\CT_{j}$ excluding the cache profiles $\ell$ where $\Bp_{\ell,g}$ is empty in round $g$. Then in round $g$, for each such subset $\CT_{j} \subseteq [L]$ 
where $\CT^g_{j}$ is nonempty\footnote{Notice that there are $\binom{L}{\gamma L+1}-\binom{v(g)}{\gamma L+1}$ number of $\CT^g$ that is nonempty where $v(g)$ is the number of empty partitions $\Bp_{\ell,g}$, for $l\in[L]$.}
, the following vector is transmitted  
\begin{equation}
\label{eq:new_transmission}
    \Bx(\CT_{j})=\sum\nolimits_{\ell \in \CT_{j}^g} z(\BH_{(\Ba_{\ell,g},\Bp_{\ell,g})}^{-1}\Bw_{(\Bp_{\ell,g})})
\end{equation}
where $\BH_{(\Ba_{\ell,g},\Bp_{\ell,g})}^{-1}$ denotes the inverse of the channel matrix between the helpers in $\Ba_{\ell,g}$ and the users in $\Bp_{\ell,g}$, 
$\Bw_{(\Bp_{\ell,g})}$ denotes the column vector where its $k$th element is   $W^{d_{\Bp_{\ell,g}[k]}}_{\CT_{j} \backslash \{ \ell \}}$ for $k\in[U_{\ell,g}]$, 
and $z$ denotes the zero-padding function that adds zeros in the helper indexes not in $\Ba_{\ell,g}$ to the vector in its argument. For instance, if $\Ba_{\ell,g}=[e_2,e_4]$, then $\BH_{(\Ba_{\ell,g},\Bp_{\ell,g})}^{-1}\Bw_{(\Bp_{\ell,g})}$ is a $2\times1$ vector, say $[V_2,V_4]^T$. Then $z([V_2,V_4]^T)=[0,V_2,0,V_4]^T$. 
\begin{exmp}
    Consider the subnetwork in Figure~\ref{fig:subnetwork}, and without loss of generality, assume that the subnetwork corresponds to users with cache profile $1$, $L=3$, $t=1$. For simplicity, we only write the indices of the users and helpers in the partition vectors, and if a partition is full, 
    we suppress the helper vector in the notation. For the second round, we have $\Bp_{1,2}=[2,5,8,12]$, assume also $\Bp_{2,2}=[14,18]$ where $\Ba_{2,2}=[2,4]$ and $\Bp_{3,2}$ is empty. There are $3$ subsets of $[L]=[3]$ of size $t+1=2$: $\CT_1=\lbrace1,2\rbrace$, $\CT_2=\lbrace1,3\rbrace$, and $\CT_3=\lbrace2,3\rbrace$. Then, in the second round, $\CT_1^2=\lbrace1,2\rbrace$, $\CT_2^2=\lbrace1\rbrace$ and $\CT_3^2=\lbrace2\rbrace$ such that $\Bx(\lbrace1,2\rbrace)=\BH^{-1}_{([2,5,8,12])}[W^{d_{u_2}}_{\lbrace2\rbrace},W^{d_{u_5}}_{\lbrace2\rbrace},W^{d_{u_8}}_{\lbrace2\rbrace},W^{d_{u_{12}}}_{\lbrace2\rbrace}]^T+ z(\BH^{-1}_{([2,4],[14,18])}[W^{d_{u_{14}}}_{\lbrace1\rbrace},W^{d_{u_{18}}}_{\lbrace1\rbrace}]^T)$, $\Bx(\lbrace1,3\rbrace)=\BH^{-1}_{([2,5,8,12])}[W^{d_{u_2}}_{\lbrace3\rbrace},W^{d_{u_5}}_{\lbrace3\rbrace},W^{d_{u_8}}_{\lbrace3\rbrace},W^{d_{u_{12}}}_{\lbrace3\rbrace}]^T$, $\Bx(\lbrace2,3\rbrace)=z(\BH^{-1}_{([2,4],[14,18])}[W^{d_{u_{14}}}_{\lbrace3\rbrace},W^{d_{u_{18}}}_{\lbrace3\rbrace}]^T)$. User $u_{14}$ can remove $\Bh_{14}^T\BH^{-1}_{([2,5,8,12])}[W^{d_{u_2}}_{\lbrace2\rbrace},W^{d_{u_5}}_{\lbrace2\rbrace},W^{d_{u_8}}_{\lbrace2\rbrace},W^{d_{u_{12}}}_{\lbrace2\rbrace}]^T$ since it has cache profile $2$, and decode its desired subfile $W^{d_{u_{14}}}_{\lbrace1\rbrace}$, as $\Bh_{14}^Tz(\BH^{-1}_{([2,4],[14,18])}[W^{d_{u_{14}}}_{\lbrace1\rbrace},W^{d_{u_{18}}}_{\lbrace1\rbrace}]^T)=W^{d_{u_{14}}}_{\lbrace1\rbrace}$. Similarly, $\Bh_{14}^Tz(\BH^{-1}_{([2,4],[14,18])}[W^{d_{u_{14}}}_{\lbrace3\rbrace},W^{d_{u_{18}}}_{\lbrace3\rbrace}]^T)=W^{d_{u_{14}}}_{\lbrace3\rbrace}$ such that $u_{14}$ can decode its requested file $W^{d_{u_{14}}}$. 
    In a similar way, all the other users can decode their requested files.
\end{exmp}
%
%
\section{Numerical Results}
\label{sec:results}

We provide simulations 
to compare the performance of the least cost branch and bound and its greedy approximation (see Example~\ref{ex:new_transmission}) and the scalable heuristic of~\cite{kagan_1} that applies to the collision interference model. As a quick explanation, the heuristic in~\cite{kagan_1} categorizes users based on the number of helpers within their transmission radius and assigns them to helpers in a greedy way by avoiding collisions.
For the simulation setup, we assume $E$ helpers are located at the center of hexagons on a limited hexagonal grid. Every hexagon has a radius of~$1$ (normalized length unit).
The users are placed in a circular area with a radius of $r_u=2.7$ centered on the center of the hexagonal lattice according to a homogeneous Poisson Point Process, 
with density per unit area $u$ (see Figure~\ref{fig:hexagonal lattice}). 
Users are randomly assigned to a cache profile $l \in [L]$. 
The error bars in the figures correspond to the standard deviation.

In Figure~\ref{fig:FixedU/L}, we plot the achievable sum-DoF as a function of $L$. 
We can observe that the least cost branch and bound method and its greedy approximation outperform the heuristic of~\cite{kagan_1} as expected, since it assumes a more restricted model, i.e., the collision interference model. 
Since $u/L$ is fixed, on average, as $L$ changes, the number of users with the same cache profile doesn't change, so, for the proposed methods, the number of partitions per cache profile remains constant. And since $\gamma L$ increases linearly as $L$ increases, the number of partitions that can be served simultaneously increases linearly, and as a consequence, we observe a linear increase in the sum-DoF.
In Figure~\ref{fig:ChangeR}, we plot the achievable sum-DoF as a function of $r$ 
together with the optimum sum-DoF
under uncoded cache placement for fully connected networks with $C_{\ell}>=E$ for all $\ell\in[L]$~\cite{parinello}. 
We can observe that as $r$ increases, the performance of the proposed methods improves, while the performance of the heuristics of~\cite{kagan_1} decreases. The reasoning is as follows: The proposed methods use the spatial multiplexing gain of the transmitters to serve multiple users by finding nonzero links between the transmitters and users. As $r$ increases, the network's connectivity increases, and sum-DoF increases. However, the heuristic of~\cite{kagan_1} serves the users by avoiding collisions, and there are fewer possibilities for collision-free transmissions. Notice that the achieved sum-DoF of the proposed methods are the same when the network becomes fully connected at $r=4.2$ as expected. 
We also see from Figure~\ref{fig:ChangeR} that the 
proposed methods achieve a sum-DoF close to optimal when the network becomes fully connected.









\bibliographystyle{IEEEtran}
\bibliography{references}

\end{document}